\documentclass[a4paper]{PoS}
\font\bigastfont=cmr10 scaled \magstep 1
\newcommand{\bdot}{\hbox{\bigastfont .}}

\newcommand{\CD}{{\cal D}}
\newcommand{\CR}{{R}}
\newcommand{\CW}{{W}}
\newcommand{\CQ}{{Q}}
\newcommand{\CS}{{S}}
\newcommand{\average}[1]{\left\langle #1 \right\rangle_\CD}

\newcommand{\laverage}[1]{\left\langle #1 \right\rangle_{\CD_{\rm \bf i}}}

\newcommand{\initial}[1]{{#1_{\rm \bf i}}}

\title{Is Dark Energy Simulated by Structure Formation in the Universe ?}

\ShortTitle{Dark Energy from Structure Formation}

\author{\speaker{Thomas Buchert}
\thanks{This work is part of a project that has received funding from the European Research Council (ERC) under the European Union's Horizon 2020 research and innovation programme (grant agreement ERC advanced Grant No. 740021--ARThUs, PI: TB). I would like to thank the organizers of the \textit{2nd World Summit: Exploring the Dark Side of the Universe}, and the \textit{Regional Council of Guadeloupe} for generous support.}\\
        Univ Lyon, Ens de Lyon, Univ Lyon1, \\CNRS, Centre de Recherche Astrophysique de
  Lyon UMR5574, F--69007, Lyon, France \\
        E-mail: \email{buchert@ens--lyon.fr}}

\abstract{
The standard model of cosmology assumes that the Universe can be described to hover around
a homogeneous--isotropic solution of Einstein's general theory of relativity.  
This description needs (sometimes hidden) hypotheses that restrict the generality, and relaxing these restrictions is the headline of 
a new physical approach to cosmology that refurnishes the cosmological framework.
Considering a homogeneous geometry as a template geometry
for the in reality highly inhomogeneous Universe must be considered a strong idealization. 
Unveiling the limitations of the standard model opens the door to rich consequences of general relativity, giving rise to 
effective (i.e. spatially averaged) cosmological models that may even explain the longstanding problems of dark energy and dark matter. \\
We explore in this talk the influence of structure formation on average properties of the Universe by discussing: (i) general thoughts on why considering average properties, on the key-issue of non-conserved curvature, and on the global gravitational instability of the standard model of cosmology; (ii) the general set of cosmological equations arising from averaging the scalar parts of Einstein's equations, the generic property of structure formation interacting with the average properties of the Universe in a scale-dependent way, and the description of cosmological backreaction in terms of an effective scalar field.
}

\FullConference{2nd World Summit: Exploring the Dark Side of the Universe\\
		25-29 June, 2018\\
		University of Antilles, Pointe-\`a-Pitre, Guadeloupe, France}

\begin{document}

\section{General Thoughts}
\label{sec:thoughts}

Before we discuss a more general cosmological framework within which we can understand inhomogeneity effects affecting global properties of world models, so-called \emph{backreaction effects}, we put our heads together on a couple of general thoughts that lie at the basis of this framework.\footnote{The reader is directed to a number of reviews on details of the presented framework \cite{Buchert08status,Buchert11Towards,BuchRas12,Buchert:rebuttal},
where the latest has especially addressed criticisms of this approach. (Further reviews by other colleagues will be cited in the text.)}

\subsection{Why averaging ?}

Cosmology is built on models for the evolution of space. Within the four-dimensional framework of general relativity we need to split the equations into spatial variables that evolve in a time-direction. Having set up such a foliation of space-time, we may select one that admits a global cosmological time, which labels spatial hypersurfaces to define a cosmological model \cite{Buchert08status,foliation:letter}.
As we observe structures in the Universe we look along the light-cone implying the need to relate observables to spatial distributions, in order to interpret them within a cosmological model. 

Within the standard model of cosmology, spatial fluctuations are conceived to evolve on an assumed background geometry that belongs to the class of 
FLRW (Friedmann--Lema\^\i tre--Robertson--Walker) cosmologies being homogeneous-isotropic solutions of Einstein's equations.
However, the description of fluctuations only makes sense with respect to their spatial average distribution and its evolution: employing the standard model we implicitly assume that it reliably describes the evolution of the universal average. However, by deriving the spatial average distribution, which is homogeneous by construction and may be large-scale isotropic to comply with constraints from the observable Universe, we cannot confirm this hypothesis: \emph{a priori}, the average does, in general, not evolve as a FLRW model. To address this issue, we have to average an inhomogeneous model on some scale, as general as possible, insisting on a background-free approach. After establishing the (scale-dependent) general averages (of evolution and constraint equations for matter \textit{but also for geometrical variables}), we may consider the average model as a (scale-dependent) `physical background model' \cite{Rocky11} with respect to which it makes sense to describe fluctuations. We shall recall this approach as it was discussed in the talk.

\subsection{Non-conservation of curvature}

The standard $\Lambda$CDM model (Cold Dark Matter model with dark energy modeled by a cosmological constant), or
\emph{concordance model}, assumes zero curvature throughout the evolution of the Universe, with structure formation being independently modeled on this homogeneous geometry. But, according to Einstein's equations, there is a geometrical side to structure formation!
Inhomogeneities in matter are coupled to inhomogeneities in geometry. Fixing a geometry and especially exactly demanding vanishing curvature may turn out to be the biggest mistake of contemporary cosmology. Why? Taking spatial curvature into account one may be mistakenly guided by the density distribution: we may think of an almost homogeneous density of `dust' (pressure-free matter) with small perturbations of some background density at some early time after recombination. Due to structure formation as a result of local gravitational instability, the rest-mass contained in the background is reshuffled into high-density objects with large voids remaining. This highly inhomogeneous density distribution of the present-day Universe averages out to the background density that was hitherto assumed, thanks to rest-mass conservation. 

We may now naively think that the same happens to the curvature distribution, i.e. an almost or exactly zero-curvature is reshuffled into a positive spatial curvature for the high-density objects \cite{estimate},
and a negative spatial curvature for the voids, but overall the distribution averages out to the almost or exactly zero curvature assumed at the beginning.  
That this is not so is a subtle issue resulting from the fact that intrinsic curvature does not obey a conservation law. 
 
Already within the class of FLRW geometries, a more realistic description is possible by including curvature: although curvature has to be constant in space, it evolves with the scale-factor of the homogeneous-isotropic model as $\propto a^{-2}$, becoming more important than the (homogeneous) density $\propto a^{-3}$ today (but less important than the contribution of a cosmological constant). This class of models has been abandoned due to tight observational constraints on the curvature at the epoch of the Cosmic Microwave Background,
\emph{for simplicity} put to zero, so that in the class of FLRW cosmologies curvature remains zero throughout the evolution. However, even if curvature is not neglected initially, the constant-curvature in the FLRW models obeys a conservation law, $k a^2 = const.$, with constant $k$, in contrast to the averaged inhomogeneous curvature that evolves differently, as we shall detail below.

Structure formation implies that the present-day Universe is volume-dominated by voids and, as a result of the 
non-conservation of curvature, it is characterized by on average negative curvature, unless the cosmological constant is assumed to exactly compensate the curvature (which is only possible on one scale) \cite{buchertcarfora}. When putting the cosmological constant to zero, the \emph{coincidence problem} that dark energy becomes relevant at the onset of nonlinear structure formation is explained due to a change of the average curvature at this epoch that evolves more strongly than a FLRW constant negative curvature model according to current estimates and modeling results, e.g. \cite{buchertcarfora,multiscale,Bolejko17}.
Even for exactly vanishing average curvature at the epoch of the Cosmic Microwave Background, intrinsic curvature emerges \cite{Buch00scalav}, and becomes on average \textit{negative} at the present epoch. This `curvature energy' qualitatively replaces dark energy in the universal energy budget. `Curvature dark energy' is of physical origin, and it is \textit{time-} and \textit{scale-dependent}, which has the potential to resolve the currently surfacing tension between observations of the Hubble expansion at different epochs and on different scales.

\subsection{Global gravitational instability of the standard model}

From first principles, even before we invoke theories of gravitation, we can show that a homogeneous-isotropic distribution pre-assumed as the `background model' at some initial time does not provide the average distribution for all times. Worse, the average distribution evolves away from this background model: there is not only the well-known local gravitational instability, there is a global one \cite{Roy:instability}. 

The \emph{averaging problem} was already raised in the 80's by George Ellis \cite{Ellis:averaging}, {\it cf.} \cite{Ellis:inhomogeneity}: we cannot expect that the inhomogeneous Universe follows---on average---a homogeneous solution of Einstein's equations. 
The cosmological framework discussed in the next section is based on a realization of George's idea, namely a realization of average properties of Einstein's equations by restricting attention to scalar variables, where averaging can be unambiguously 
implemented through Riemannian volume averages of the \emph{general set} (i.e. without any symmetry assumption or approximation) of the scalar parts of Einstein's equations within a $3+1$ splitting of spacetime \cite{Buch00scalav}. The result recovers the cosmological equations of Friedmann and Lema\^\i tre, however, with additional source terms---so--called backreaction terms---that can mimic dark energy on large scales, but they can also add to an effect similar to the existence of dark matter by going to smaller scales. 

The key--reason for the existence of additional terms that encode geometrical inhomogeneities is the \emph{non--commutation of the averaging process and (rest-mass preserving) evolution}, even before we specify a theory of gravitation \cite{Buch00scalav,ellisbuchert}: the averaging process is \emph{non-local} and implies that the evolution of an averaged distribution is not the average of the evolution of the local distribution. To show this, we define
volume averaging, restricting 
attention to scalar functions $\Psi (t,X^i )$,
\begin{equation}
\label{eq:average-GR}
\average{\Psi (t, X^i)}:=
\frac{1}{V_\CD}\int_\CD \Psi (t, X^i)\;J \mathrm{d}^3 X \;\;;\;\;V_\CD = \int_\CD J \mathrm{d}^3 X \;\;\;\; \Rightarrow \;\;\average{\Theta} = \frac{{\dot V}_\CD}{V_\CD} \;,
\end{equation}
with $J:=\sqrt{\det(g_{ij})}$, $\Theta := \dot J / J$ being the local expansion rate, $g_{ij}$ an arbitrary metric  of the
spatial  hypersurfaces, and  $X^i$ local coordinates that  are constant
along the flow congruence. We restrict our attention in what follows to a simple matter model (`irrotational dust'), where averaging is
over a collection of free--falling fluid elements contained within a simply--connected domain 
$\CD$ with volume $V_\CD = |\CD |$ in a t--hypersurface, which is \emph{comoving} (hence, there is no issue of matter flow across the boundary of the averaging domain). These hypersurfaces are defined by constant proper time of all fluid elements, providing a synchronous foliation of space-time and a global cosmological time.
Following from (\ref{eq:average-GR}), we obtain
the {\it commutation rule} for a scalar function $\Psi$:
\begin{equation}
\label{commutationrule}
\langle{\Psi}\dot{\rangle}_\CD - \langle{\dot\Psi}\rangle_\CD = \average{\Theta\Psi} - \average{\Theta}\average{\Psi}\;\;.
\end{equation}
We apply this rule to two scalars relevant for cosmology; we obtain for the expansion rate, $\Psi = \Theta$,
\begin{equation}
\label{expansion}
\langle{\Theta}\dot{\rangle}_\CD - \langle{\dot\Theta}\rangle_\CD = \average{\Theta^2} - \average{\Theta}^2 =\average{\left(\Theta - \average{\Theta}\right)^2}\;,
\end{equation}
and for the density $\Psi = \varrho$, its deviation from the average, $\delta\varrho : = \varrho - \average{\varrho}$, and the corresponding deviation for the expansion rate, $\delta\Theta : = \Theta - \average{\Theta}$,
\begin{equation}
\label{entropy}
\langle{\varrho}\dot{\rangle}_\CD - \langle{\dot\varrho}\rangle_\CD = \average{\Theta\varrho} - \average{\Theta}\average{\varrho} = \frac{\dot{\CS}_\CD}{V_\CD} = \average{\delta\varrho\delta\Theta}\;,\;\;\;{\rm with} \quad \CS_\CD : = \int_\CD \frac{\varrho - \average{\varrho}}{\average{\varrho}} \; J \mathrm{d}^3 X \;. 
\end{equation}
Firstly, let us look at the volume expansion rate, averaged on different spatial scales. While the FLRW cosmologies are blind to a scale--dependent expansion rate, the averaged equations allow us to describe differential expansion: expansion slows down significantly on the scale of virialized structures, so that, as a result of this, large expansion variance between these small scales and the scales of voids and beyond is produced. This leads to a non-local repulsive effect in the averaged equations counteracting gravitation, a similar effect as that of a positive cosmological constant.
This can result in volume acceleration \cite{Buchert08status,multiscale,Rasanen:directions,Buchert11Towards,BuchRas12}, despite the fact that all fluid elements within the averaging domain decelerate if we assume vanishing vorticity and a negative or zero cosmological constant \cite{Raychaudhuri}. 

Secondly, the source for non-commutation in the case of the density is given by the volume-weighted production of a functional that is known in information theory as the Kullback-Leibler relative entropy $\CS_\CD$ \cite{hosoya:entropy}. Looking at the last equality in (\ref{entropy}), we infer that this entropy grows for contracting overdensities ($\delta\varrho > 0, \delta\Theta <0$) and for expanding underdensities ($\delta\varrho < 0, \delta\Theta > 0$). These two states are favoured by gravity (inhomogeneities are enhanced; gravity provides a negative feedback in the \textit{open} system contained in the domain $\CD$).

Both of the considerations above lead to the same conclusion: the average model is driven away from the FLRW solution (with vanishing expansion variance and vanishing relative entropy). 

\section{General Cosmological Equations}

We are going to work with Einstein's equations, restricted to irrotational fluid motion
with the matter model `dust' (i.e. vanishing pressure).\footnote{The corresponding average equations for a perfect fluid 
for vanishing shift in a flow-orthogonal foliation are discussed in \cite{Buch01scalav},
for general fluids in arbitrary foliations see \cite{foliation:letter}. Note that the averaging operation will produce an effective (geometric) pressure in the effective energy-momentum tensor, even if we consider a `dust' source.}

We employ a foliation of  spacetime with the 3--metric coefficients $g_{ij}$ in the $4-$metric form
\begin{equation}
{}^4 {\bf g} = -\mathrm{d}t^2 + {^3}{\bf g}\;\;\;;\;\;\;{}^3 {\bf g} = g_{ij}\,{\mathrm d}X^i \otimes {\mathrm d}X^j\;\;.
\end{equation} 
For irrotational dust the flow is geodesic and space--like hypersurfaces can be
constructed that are flow--orthogonal at every spacetime event in a $3+1$ representation. 

\subsection{Friedmann's equations versus general cosmological equations}

Friedmann's differential equations capture the scalar parts of Einstein's equations,
while subjecting them to the strong symmetry assumption of \emph{local isotropy}, hence to a homogeneous metric form with spatial coefficients $g_{ij}^F = a^2(t) k_{ij}$, with a constant curvature metric $k_{ij}$.
The resulting equations, Friedmann's expansion law (the energy constraint) and Friedmann's acceleration law (Raychaudhuri's equation), together with rest-mass conservation,
\begin{equation}
\label{friedmann}
3\left(\frac{\dot a}{a}\right)^2 - 8\pi G\varrho_H  - \Lambda +\frac{3k}{a^2}\;=\;0\quad;\quad
3\frac{\ddot a}{a} + 4\pi G\varrho_H - \Lambda\;=\;0\quad;\quad
\dot \varrho_H + 3\frac{\dot a}{a} \;\varrho_H = 0 \;,
\end{equation}
can be replaced by their spatially averaged, {\em general} counterparts for an arbitrary $3-$metric $g_{ij}$ \cite{Buch00scalav}):
\begin{eqnarray}
\label{averagedhamilton}
3\left( \frac{{\dot a}_\CD}{a_\CD}\right)^2 - 8\pi G \average{\varrho}-\Lambda +\frac{3k_{\initial\CD}}{a_\CD^2}\;=\; - \frac{1}{2}\left(\CW_\CD+\CQ_\CD\right) \;; \\
\label{averagedraychaudhuri}
3\frac{{\ddot a}_\CD}{a_\CD} + 4\pi G \average{\varrho} -\Lambda\;=\; {\CQ}_\CD\;\;; \\
\label{averagedcontinuity}
\langle{\varrho}\rangle\dot{}_\CD + 3\frac{{\dot a}_\CD}{a_\CD} \average{\varrho}\;=\;0\;\;.
\end{eqnarray}
We have replaced the Friedmannian scale factor $a(t)$ by the \emph{volume scale factor} $a_\CD (t)$, depending  on content, shape and position of the domain of averaging $\CD$,  defined via the
domain's volume  $V_{\CD}(t)=|\CD|$, and the initial 
volume $V_{\initial\CD}=V_{\CD}(\initial{t})=|\initial{\CD}|$:\footnote{Using a scale factor instead of the volume should not be confused with `isotropy'. The above
equations are general for the evolution of a mass-preserving, compact domain containing
an irrotational continuum of `dust', i.e. they
provide a background-free and non-perturbative description of inhomogeneous \emph{and} anisotropic fields.}
\begin{equation}
\label{volumescalefactor}
a_\CD (t) := \left( \frac{V_{\CD}(t)}{V_{\initial\CD}} \right)^{1/3} \;\;.
\end{equation}
The new term appearing in these equations, the {\em kinematical backreaction} ${\CQ}_\CD$, arises as
 a result of expansion and shear inhomogeneities:\footnote{$\rm I$  and $\rm II$  denote  the principal scalar invariants  of the expansion tensor, and the second equality follows by introducing the decomposition
of the expansion tensor into its kinematic parts, $\Theta_{ij} = \frac{1}{3} \Theta g_{ij} + \sigma_{ij}$, with the rate of expansion $\Theta$ and the shear $\sigma_{ij}$. The overdot denotes covariant time-derivative.
}
\begin{equation}
\label{backreactionterm} 
{\CQ}_\CD : = 2 \average{\rm II} - \frac{2}{3}\average{\rm I}^2 =
\frac{2}{3}\average{\left(\Theta - \average{\Theta}\right)^2 } - 
2\average{\sigma^2} \;.
\end{equation}
The general averaged 3--Ricci curvature $\average{\CR}$ replaces the constant--curvature term in Friedmann's equations; it is written above as a new backreaction variable that captures the deviation of the average scalar curvature from a (scale-dependent) constant curvature model, $\CW_\CD := \average{\CR} - 6k_\initial\CD / a^2_\CD$.

\subsection{Structures `talk' to the background}

In the Friedmannian case, Eqs.~(\ref{friedmann}), the acceleration law arises as the time--derivative
of the expansion law, if the integrability condition of rest-mass conservation is respected,
i.e. the homogeneous density $\varrho_H \propto a^{-3}$. In the general case, however, 
rest-mass conservation is not sufficient. In addition to the general integral of Eq.~(\ref{averagedcontinuity}), 
$\average{\varrho} = \laverage{\varrho(\initial{t})} / a_\CD^3$, 
we also have to respect the following {\em curvature--fluctuation--coupling} furnishing a new conservation law \cite{Buch00scalav}:
\begin{equation}
\label{integrability}
\frac{1}{a_\CD^{2}} \;\left(\,\CW_\CD \,a_\CD^2 \,\right)^{\bdot}\;= - \frac{1}{a_\CD^6}\left(\,{\CQ}_\CD \,a_\CD^6 \,\right)^{\bdot} \;\;.
\end{equation}
This integrability condition assures that (\ref{averagedraychaudhuri}) is the time-derivative of  (\ref{averagedhamilton}). It belongs to the general cosmological equations and relates intrinsic and extrinsic curvature invariants, and it is key to understand how backreaction can take the role of dark energy. Although the above description of inhomogeneities is background-free, we are entitled to advance the 
point of view to consider $a_\CD (t)$ as the scale--factor of a `background', defined by the averaged dynamics, that in general couples to the fluctuations. A global instability that drives the averaged model away from a Friedmannian solution has been identified for $\lbrace\CQ_\CD > 0\;,\;\CW_\CD <0 \rbrace$ (on scales dominated by differential expansion and leading to dark energy-like behavior), and for
$\lbrace\CQ_\CD < 0\;, \CW_\CD > 0\rbrace$ (on scales dominated by shear and leading to dark matter-like behavior)   \cite{Roy:instability}.

In the standard model the fluctuations `do not talk' to the background model universe that remains unaltered, while in reality the average model universe is driven away from a prescribed background as a result of the interaction with the formation of structure \cite{Roy:instability}.\footnote{Recall that in the standard model, structure formation is described to separately evolve on the background model. While the background model `talks to structure formation' in the sense that structures have a harder time to form if the background expansion is faster, the opposite crosstalk of structure formation on the background model is suppressed.} 

We may recast the general equations (\ref{averagedhamilton}, \ref{averagedraychaudhuri}, \ref{averagedcontinuity}, \ref{integrability}) by appealing to the Friedmannian framework.
This amounts to re--interpret averaged-out geometrical terms as 
effective {\em sources} within a Friedmannian setting. This leads to scale-dependent
Friedmann equations for an {\em effective perfect fluid energy momentum tensor}
with new effective sources \cite{Buch00scalav,Buch01scalav}:
\begin{eqnarray}
\label{effectivesources}
&\varrho^{\CD}_{\rm eff} := \average{\varrho}-\frac{1}{16\pi G}{\CQ}_\CD - 
\frac{1}{16\pi G}\CW_\CD\;\;\;;\;\;\;
{p}^{\CD}_{\rm eff} :=  -\frac{1}{16\pi G}{\CQ}_\CD + \frac{1}{48\pi G}\CW_\CD\;\;.\\
\label{effectivefriedmann}
&3\left(\frac{{\dot a}_\CD}{a_\CD}\right)^2 - 8\pi G \varrho^{\CD}_{\rm eff} - \Lambda+\frac{3k_\CD}{a_\CD^2}\;=\;0\;\;;\;\;
3\frac{{\ddot a}_\CD}{a_\CD}  +  4\pi G (\varrho^{\CD}_{\rm eff}
+3{p}^{\CD}_{\rm eff}) - \Lambda \;=\;0\;\;;\nonumber\\
&{\dot\varrho}^{\CD}_{\rm eff} + 
3 \frac{{\dot a}_\CD}{a_\CD} \left(\varrho^{\CD}_{\rm eff}
+{p}^{\CD}_{\rm eff} \right)\;=\;0\;\;.
\end{eqnarray}
Eqs.~(\ref{effectivefriedmann}) correspond to the equations
(\ref{averagedhamilton}),
(\ref{averagedraychaudhuri}), (\ref{averagedcontinuity}) and (\ref{integrability}), respectively. 

\subsection{General cosmological equations as `morphed' Friedmann cosmologies}

We notice from (\ref{effectivesources}) that ${\CQ}_\CD$, if interpreted as a source, introduces a component with `stiff equation of state', $p^\CD_{\CQ} = \varrho^\CD_\CQ$, suggesting a correspondence with
a free scalar field, while the averaged scalar curvature introduces a component with `curvature equation of state'
$p^\CD_{\CW} = -1/3 \varrho^\CD_{\CW}$. Although we are dealing with `dust', we appreciate a `geometrical pressure' in the effective energy-momentum tensor composed of $\varrho_{\rm eff}  =: \average{\varrho} + \varrho_{\psi}$
and $p_{\rm eff}  = : p_{\psi}$, where the \emph{backreaction fluid} can be accordingly written in scalar field language 
as follows \cite{morphon}:
	\begin{equation}
	\label{fluid_morph}
\varrho_{\psi} = \epsilon \frac{1}{2}\dot{\psi}_\CD^2 + U_\CD\left(\psi_\CD\right)\quad;\quad
		p_{\psi}  = \epsilon \frac{1}{2}\dot{\psi}_\CD^2 - U_\CD\left(\psi_\CD\right)\;.
\end{equation}
The dictionary that translates physical variables to scalar field variables reads:
\begin{equation}
	\dot{\psi}^2_\CD = -\epsilon\frac{\CW_\CD + 3{\CQ_\CD}}{24\pi G}\quad;\quad
	U_\CD = \frac{-\CW_\CD}{24\pi G}\;,
	 \label{eq:U}
\end{equation}
where the sign of $(\CW_\CD + 3{\CQ_\CD})(t)$ determines the transition from a real ($\epsilon = +1$) to a phantom ($\epsilon = -1$) field. The \emph{morphon} is effective so that there is no violation of energy conditions at this transition point. The effective conservation equation (the third equation of (\ref{effectivefriedmann})) becomes the \emph{Klein-Gordon equation} for the scalar field $\Psi_\CD$:
\begin{equation}
		 \ddot{\Psi}_\CD + 3H\dot{\Psi}_\CD + \epsilon \frac{\mathrm{d} U_\CD (\Psi_\CD)}{\mathrm{d} \Psi_\CD}\;=\;0\;. \label{eq::KG_D}
\end{equation}
The system of equations~(\ref{effectivefriedmann}) written in terms of the morphon field is formally equivalent to the Friedmann equations describing a model universe sourced by dust-matter and a fundamental, minimally coupled scalar field evolving in a given potential.

\section{Where do we go from here ?}

From what we discussed it is evident that the standard model of cosmology suffers from omissions of physics in a general-relativistic context, most prominently, global curvature evolution is neglected in the so-called concordance cosmology. A background-free and fully relativistic approach shows that a physical cosmological model must account for the evolution of global properties of universe models due to the interaction of its average properties with fluctuations in matter and geometry.
Whether cosmological backreaction can fully replace the need for dark energy and even dark matter is an open question that needs generic models whose architecture is not restricted as in Newtonian cosmology \cite{buchert:newtonian}.
Since observations are interpreted within the standard model, the quantitative investigation of inhomogeneous models has to be accompanied by a re-interpretation of observational data in these new models to reach a conclusive answer.

\end{document}